\documentclass[twocolumn]{aastex62}
\graphicspath{{./}{figures/}}
\pdfoutput=1
\received{October 18, 2018}
\revised{November 8, 2018}
\accepted{November 9, 2018}
\submitjournal{\apj}
\shorttitle{HD 156623 $\delta$ Scuti Pulsations}
\shortauthors{Mellon et al.}

\begin{document}

\title{Discovery of $\delta$ Scuti Pulsations in 
the Young Hybrid Debris Disk Star HD 156623}

\correspondingauthor{Samuel N. Mellon}
\email{smellon@ur.rochester.edu}
\author[0000-0003-3405-2864]{Samuel N. Mellon}
\affil{Department of Physics \& Astronomy, University of Rochester, 500 Wilson Blvd., Rochester, NY 14627, USA}

\author[0000-0003-2008-1488]{Eric E. Mamajek}
\affil{Jet Propulsion Laboratory, California Institute of Technology, M/S 321-100, 4800 Oak Grove Drive, Pasadena, CA 91109, USA}
\affil{Department of Physics \& Astronomy, University of Rochester, 500 Wilson Blvd., Rochester, NY 14627, USA}

\author[0000-0001-9229-8315]{Konstanze Zwintz}
\affil{Institut f{\"u}r Astro- und Teilchenphysik, Universit{\"a}t Innsbruck, Technikerstrasse 25/8, 6020 Innsbruck, Austria}

\author[0000-0001-6534-6246]{Trevor J. David}
\affil{Jet Propulsion Laboratory, California Institute of Technology, M/S 321-100, 4800 Oak Grove Drive, Pasadena, CA 91109, USA}

\author[0000-0001-7797-3749]{Remko Stuik}
\affil{Leiden Observatory, Leiden University, PO Box 9513, 2300 RA Leiden, The Netherlands}

\author[0000-0003-4787-2335]{Geert Jan J. Talens}
\affil{Leiden Observatory, Leiden University, PO Box 9513, 2300 RA Leiden, The Netherlands}

\author[0000-0003-3812-2436]{Patrick Dorval}
\affil{Leiden Observatory, Leiden University, PO Box 9513, 2300 RA Leiden, The Netherlands}

\author[0000-0002-2487-4533]{Olivier Burggraaff}
\affil{Leiden Observatory, Leiden University, PO Box 9513, 2300 RA Leiden, The Netherlands}
\affil{Institute of Environmental Sciences (CML), Leiden University, PO Box 9518, 2300 RA Leiden, The Netherlands}

\author[0000-0002-7064-8270]{Matthew A. Kenworthy}
\affil{Leiden Observatory, Leiden University, PO Box 9513, 2300 RA Leiden, The Netherlands}

\author[0000-0002-4272-263X]{John I. Bailey, III}
\affil{Department of Physics, University of California at Santa Barbara, Santa Barbara, CA 93106, USA}

\author[0000-0002-1520-7851]{Blaine B. D. Lomberg}
\affil{South African Astronomical Observatory, Observatory Rd, Observatory Cape Town, 7700 Cape Town, South Africa}
\affil{Department of Astronomy, University of Cape Town, Rondebosch, 7700 Cape Town, South Africa}

\author[0000-0002-4236-9020]{Rudi B. Kuhn}
\affil{South African Astronomical Observatory, Observatory Rd, Observatory Cape Town, 7700 Cape Town, South Africa}

\author[0000-0002-6194-043X]{Michael J. Ireland}
\affil{Research School of Astronomy and Astrophysics, Australian National University, Canberra, ACT 2611, Australia}

\author[0000-0002-8969-5229]{Steven M. Crawford}
\affil{South African Astronomical Observatory, Observatory Rd, Observatory Cape Town, 7700 Cape Town, South Africa}
\affil{Space Telescope Science Institute, 3700 San Martin Drive, Baltimore, MD 21218, USA}

\begin{abstract}
The bRing robotic observatory network was built to search for circumplanetary material within the 
transiting Hill sphere of the exoplanet $\beta$ Pic b across its bright host star $\beta$ Pic.
During the bRing survey of $\beta$ Pic, it simultaneously monitored the brightnesses of thousands 
of bright stars in the southern sky ($V$ $\simeq$ 4-8, $\delta$ $\lesssim$ -30$^{\circ}$). 
In this work, we announce the discovery of $\delta$ Scuti pulsations in the A-type star HD 156623 
using bRing data. HD 156623 is notable as it is a well-studied young star 
with a dusty and gas-rich debris disk, previously detected using ALMA. 
We present the observational results on the pulsation periods and amplitudes for HD 156623, discuss its evolutionary status, and provide further constraints on its nature and age. We find strong evidence of frequency regularity and grouping. We do not find evidence of frequency, amplitude, or phase modulation for any of the frequencies over the course of the observations. We show that HD 156623 is consistent with other hot and high frequency pre-MS and early ZAMS $\delta$ Scutis as predicted by theoretical models and corresponding evolutionary tracks, although we observe that HD 156623 lies hotter
than the theoretical blue edge of the classical instability strip. This, coupled with our characterization and Sco-Cen membership analyses, suggest that the 
star is most likely an outlying ZAMS member of the $\sim$16 Myr Upper Centaurus-Lupus
subgroup of the Sco-Cen association.
\end{abstract}


\section{Introduction \label{sec:intro}}

The number of ground-based, wide-field photometric surveys that have been designed to detect transiting exoplanets  has grown considerably over the past two decades \citep[e.g.,][]{Bakos02,Pollacco06,Pepper07,Talens17}. In addition to exoplanets, these surveys are sensitive to photometric variability due to eclipsing binaries, rotation periods, pulsations, amongst others \citep[e.g.,][]{Mamajek12,Mellon17,Oberst17,Burggraaff18}. These discoveries have been possible due to the high-cadence, high-precision photometry of stars typically fainter than $V \simeq 7$. Fainter stars have been targeted due to their greater abundance per frame in the sky, making these surveys highly efficient.

These observations require exposure times on the order of tens of seconds to minutes. Brighter stars have been typically excluded from these surveys due to saturation of the detectors in their exposures. The bRing ({\it Beta Pic b Ring}) survey \citep[][]{Stuik17} is a ground-based, wide-field photometric survey designed to monitor bright stars ($V$ $\simeq$ 4--8), including the bright nearby exoplanet and debris disk host star $\beta$ Pic, and observes stars brighter than many of the ground-based exoplanet surveys. Due to its high cadence and photometric precision, bRing is able to detect sub-mmag signals for these bright stars and is sensitive to large exoplanets and faint stellar signals, including $\delta$ Scuti pulsations. This paper presents the first science result from the bRing survey.

$\delta$ Scuti variable stars have been known to exhibit surface radial and non-radial pressure and gravitational pulsation modes \citep[e.g.,][]{Fitch81, Balona98, Breger00, Guenther09, Zwintz14}. 
One goal of asteroseismologists has been to develop models to identify and characterize these pulsations; such models can be used to probe the structure of a star \citep[e.g.,][]{Kurtz14} and help characterize its age and evolutionary status \citep[e.g.,][]{Guenther09,Zwintz14c}. These pulsation frequencies are also known to exhibit regular spacing in frequency and appear in groups \citep[e.g.,][]{Zwintz11,Zwintz14,Kurtz14}. $\delta$ Scutis have the tendency to lie within the classical instability strip on the HR diagram \citep{Breger98}, with some known to lie outside the strip \citep{Bowman18}.  

HD 156623 (HIP 84881, 2MASS J17205061-4525149) is a bright \citep[$V$ = 7.25; ][]{Cousins62,Mermilliod91} A1\,V star \citep{Paunzen01}.  Since 1962, it has served as a Harvard E region photometric standard in 
several systems, starting with the 
$UBV$ system \citep{Cousins62, Menzies80, Cousins83, Menzies89} and later 
the $VR_{c} I_{c}$ \citep{Cousins76, Menzies80, Cousins80, Menzies89}, 
$uvby$ \citep{Cousins87, Kilkenny92}, and
H$\beta$ systems \citep{Cousins90}. 
Reported mean photometry on the $uvby\beta$ system is provided in the compendia
of \citet{Hauck97} and \citet{Paunzen15}, and on the $UBV$ system by 
\citet{Mermilliod91}.

The star was included both as a photometric standard, and as being an A-type star in the vicinity of the Sco-Cen OB association by \citet{Slawson92} (although no membership analysis was conducted). 
The star was first noted as having an infrared excess in the WISE survey
by \citet{Mcdonald12}.
\citet{Rizzuto12} considered the star (listed as HIP 84881) as a low probability member
of Sco-Cen. 
The star's membership to Sco-Cen is not obvious as it lies well below the Galactic
plane ($b$ = -4.8$^{\circ}$) on the opposite side of the plane from the Upper Centaurus-Lupus
subgroup of Sco-Cen \citep[spanning 0$^{\circ}$ $<$ $b$ $<$ 25$^{\circ}$;][]{deZeeuw99}.
Trigonometric parallaxes of $\varpi$ = 8.45\,$\pm$\,0.60 mas \citep[][using Hipparcos astrometry]{vanLeeuwen07} and $\varpi$ = 8.9484\,$\pm$\,0.0769 mas \citep{Gaia16,GaiaDR2} have been reported, consistent with a distance $d$ = 111.41$^{+0.97}_{-0.95}$ pc 
\citep{Bailer-Jones18}.
\citet{Lieman-Sifry16} report ALMA observations of HD 156623, both 1240 $\mu$m continuum and $^{12}$CO(2-1) emission.
Although detection of gas among young debris disk stars was relatively low in Sco-Cen ($\sim$16\%) \citep{Lieman-Sifry16}, ALMA surveys focusing specifically on young A-type stars are now showing that CO gas appears to be very common
\citep{Moor17, Kral17}.

The work presented in this paper reports bRing's detection of $\delta$ Scuti pulsation frequencies in the A-type star HD 156623 and provides a characteristic analysis of the star and these frequencies. The paper is organized as follows: \S \ref{sec:observations} describes the bRing observations; \S \ref{sec:data} details the data selection, reduction, frequency extraction, and stellar characterization; \S \ref{sec:discussion} discusses the refined properties of HD 156623, reassessing its membership to Sco-Cen and estimated age, and the reported $\delta$ Scuti pulsation frequencies and their analysis.

\section{Observations \label{sec:observations}}
\subsection{bRing Photometry}

The $\beta$ Pictoris b Ring (bRing) survey was designed to monitor the 2017 -- 2018 transit of the Hill sphere of the young giant exoplanet $\beta$ Pic b in front of the bright ($V$ = 3.8) star $\beta$ Pic, and search for evidence of any circumplanetary matter \citep{Wang16, Stuik17, Mellon18}. One telescope (constructed at Leiden University) was installed in Sutherland, South Africa -- an observing facility of the South African Astronomical Observatory in Cape Town, South Africa -- and the other (constructed at University of Rochester) at Siding Spring Observatory in Coonabarabran, NSW Australia. The South African bRing observatory has been observing continuously since 2017 January. The Australian bRing observatory started in 2017 November.

Each bRing station has two FLI Microline ML11002M cameras that were oriented to maximize the longitudinal coverage of $\beta$ Pic, which is nearly circumpolar at both sites (latitudes $\simeq$ -32$^{\circ}$). One camera was pointed southeast (Az = 150$^{\circ}$) and the other southwest (Az = 210$^{\circ}$). The cameras came equipped with 4008$\times$2672 pixel Grade 2 CCDs, which have a QE of $\sim$50\% and 9\,$\mu$m pixels. The cameras were attached to Canon 24\,mm f/1.4 wide-field lenses; this resulted in a 74$^{\circ}$$\times$53$^{\circ}$ FOV with $\sim$1\arcmin\, pixels. The data were taken with alternating exposures of 6.38 and 2.54 seconds \citep{Stuik17}.

The bRing calibration pipeline builds on the heritage of the MASCARA data pipeline \citep{Stuik14,Talens17,TalensC18}. The bRing pipeline \citep[discussed in][]{Stuik17,TalensC18} utilizes relative aperture photometry calibrated to the ASCC catalog \citep{Kharchenko01}, which itself is tied to the $BV$ photometry of the {\it Hipparcos} and {\it Tycho} instruments \citep{ESA97}. The pipeline performs a preliminary data reduction correcting CCD quantum efficiency, total throughput of the lenses and sky, intra-pixel variations, and sky- and cloud-based temporal variations. An astrometric solution is calculated every 50 exposures. For each camera, the resulting data are binned to a 5 minute cadence and updated for download on a bi-weekly basis.

In addition to $\beta$ Pic, the survey also monitored the brightnesses of $\sim$20,000 of the brightest stars in the southern sky ($V$ $\simeq$ 4--8), granting it a unique role amongst ground-based, wide-field photometric surveys \citep[similar to its sibling MASCARA;][]{Snellen12,Snellen13,Stuik14,Talens17,TalensC18}. The combined sky coverage of both sites has provided nearly continuous coverage of stars (an ideal 24 hour day provides 20-21 hours of continuous coverage). This nearly continuous temporal coverage combined with a  cadence of 5 minutes has made bRing sensitive to serendipitous aperiodic and high-frequency, low amplitude periodic events. The observations of HD 156623 in this work cover 2017 June -- 2018 May.

\subsection{FEROS Spectrum}
A FEROS spectrum of HD 156623 taken from the European Southern Observatory (ESO) archive was used in this analysis. The FEROS instrument is a high-resolution (R=48000) \'Echelle spectrograph located at the ESO in La Silla, Chile \citep{Kaufer99}. Science products processed through the FEROS Data Reduction System are available for query and download on the ESO data archive\footnote{\url{http://archive.eso.org/wdb/wdb/adp/phase3\_spectral/form?collection\_name=FEROS}}.

Although HD 156623 has been observed by FEROS several times at various exposure times, we analyzed a single spectrum with the highest SNR \citep[ADP.2016-09-28T11:26:17.118;][]{Rebollido18}. This observation has the longest exposure time (1800.04 seconds) and the highest SNR (399.0). For the purposes of this work, this single spectrum was sufficient for providing estimates on the radial velocity and $v$sin$i$.

\section{Analysis \label{sec:data}}

\subsection{Data Sample, Reduction, and Noise Discussion \label{noise}}

A subset of the bRing light curves were originally chosen to search for new variable stars. Stars were chosen if they had no previously reported variability \citep[simply assessed via lacking an entry in the AAVSO VSX catalog;][]{Watson06}\footnote{The VSX catalog is regularly updated at \url{http://cdsarc.u-strasbg.fr/viz-bin/Cat?B/vsx.}} and no bright ($V$ $\lesssim$ 10) neighboring stars within a 10\arcmin\, radius (corresponding to 4 bRing inner apertures). These measurements were also used to help identify major systematics still present in the calibrated data.

Beyond the calibration, this study detrended additional strong systematics identified in the data. Due to the high cadence and stationary nature of bRing, strong systematic signals with frequencies at 1 sidereal day and 1 synodic day were generated from each star traveling across the CCD, as well as a systematic signal due to the Moon \citep{Stuik17, Burggraaff18, TalensC18}. These three sets of periodicities were simultaneously fit and detrended using an iterative median-binning routine. A second-order CCD QE response systematic was also fitted and subtracted for each camera. A heliocentric correction to the time series was also applied. Finally, the data from the individual cameras were median aligned to each other in order to remove scaling offsets due to the $\sim$1\%\, differences in the calibrations. Composite light curves containing data from all 4 cameras were then generated. The detrended rms in each camera was also calculated for comparison to the predicted theoretical noise in each star.

At the time of this publication, the principal theoretical sources of Poisson noise have been quantified to first order, i.e., shot, background, dark current, and read noise. The uniqueness of the bRing survey and the detrended rms measurements of each star imply additional higher-order considerations must be included such as non-uniform transmission. We are presently working to incorporate these sources to better understand the residual noise and improve our post-calibration detrending routine. The goals of these analyses are to provide a post-calibration detrending routine that minimizes the residual noise in each star and to determine an overall scintillation noise-floor for each camera.

\subsection{Lomb-Scargle Analysis and Frequency Extraction \label{extracting}}

The Lomb-Scargle (LS) periodogram \citep{Press92,Scargle82} is useful for detecting periodic signals in unevenly sampled data \citep[e.g.,][]{Hartman08,Hartman10,Messina10,Cargile14,Mellon17}. The LS periodogram function available in the Python {\it Scipy}\footnote{\url{https://scipy.org/}} package was used to verify the removal of the major systematics and to search for stars in this sample that show signs of previously undetected variability. For the periodograms in this work, the Nyquist frequency \citep{Press92} of $\sim$135 $d^{-1}$ was used as an upper limit on the frequency with frequency spacings of $\sim$5$\times$10$^{-4}$  $d^{-1}$. The {\it Scipy} normalized periodograms were generated for the composite light curve and four camera light curves.

The major systematics (sidereal, synodic, lunar) were clearly filtered out by our process with their peaks being cut by a factor of $\sim$100. There were a few remaining low frequency ($<$ 5 $d^{-1}$) peaks that were only a few tenths of mmags above the residual noise level. These could be residual beats, interference frequencies, or unidentified systematics in bRing. Other than one possible systematic at $\sim$24 $d^{-1}$ shared by some of the stars in this sample, there were no other systematic peaks in the periodograms of these stars. To check for new variables, these stars were studied in order of largest peak in the composite light curve periodogram until stars whose peaks were consistent with the noise level were found. This final sample of stars with peaks above the noise in their periodograms were phase-folded on their strongest periods and analyzed further to see if new variability or systematics were detected, i.e., the detected variability was either exclusive to a star or shared with other stars.

This analysis yielded evidence of $\delta$ Scuti pulsations for HD 156623, previously unrecognized as a variable likely due to its high frequency ($f_{detected} >$ 56 $d^{-1}$), low amplitude ($<$ 10 mmag) pulsations. All 4 bRing cameras had reported a total of $\sim$12500 data points between 2017 June 1 and the download date of 2018 May 28. The tool used to extract the frequencies was the literature standard {\tt Period04} software \citep{Lenz05}. The {\tt Period04} analysis was performed on the composite light curve and the 4 individual camera light curves of HD 156623. The prewhitening residual noise in the Fourier spectrum from {\tt Period04} for the composite light curve was $\sim$0.2 mmag. Periods were searched for down to SNR = 4 \citep{Kuschnig97}, which corresponded well with visual inspection of the Fourier spectrum revealing no more significant periods. The detected frequencies were entered into a spreadsheet for selection, analysis, and discussion ({\S \ref{dsfreq}}).

\begin{deluxetable}{cccc}
\tablecaption{Adopted Observational Values for HD 156623\label{tab:HD156623}}
\tablehead{
\colhead{Parameter} & \colhead{Value} & \colhead{Units} & \colhead{Reference}}
\colnumbers
\startdata
$\alpha$       & 260.21081592$\tablenotemark{a}$ ($\pm$0.05 mas) & deg & 1\\
$\delta$       & -45.42100519$\tablenotemark{a}$ ($\pm$0.06 mas) & deg & 1\\
$\varpi$       &  8.9484\,$\pm$\,0.0769 & mas & 1\\
$d$            &  111.41$^{+0.97}_{-0.95}$ & pc & 2\\
$\mu_{\alpha}$ &  -14.3\,$\pm$\,0.1 & mas\,yr$^{-1}$ & 1\\
$\mu_{\delta}$ &  -40.1\,$\pm$\,0.1 & mas\,yr$^{-1}$ & 1\\
SpT            &   A1\,V               & ... & 3\\
$V$            &   7.254\,$\pm$\,0.007 & mag & 4\\
$B-V$          &   0.087\,$\pm$\,0.003 & mag & 4\\
($b-y$)        &   0.040\,$\pm$\,0.003 & mag & 5\\
$m_1$          &   0.193\,$\pm$\,0.004 & mag & 5\\
$c_1$          &   0.960\,$\pm$\,0.002 & mag & 5\\
$\beta$        &   2.904\,$\pm$\,0.001 & mag & 5\\
E($B-V$)       &  0.008\,$\pm$\,0.018  & mag & 6\\
\enddata
\tablenotetext{a}{epoch J2015.5, ICRS}
\tablecomments{
(1) \citet{GaiaDR2},
(2) \citet{Bailer-Jones18},
(3) \citet{Paunzen01},
(4) \citet{Mermilliod91},
(5) \citet{Paunzen15},
(6) STILISM \citep{Lallement18}.
}
\vspace{-0.4in}
\end{deluxetable}

\subsection{Stellar Characteristics \label{characteristics}}

In preparation for the intended $\delta$ Scuti frequency analysis, we performed a stellar characterization analysis of HD 156623 to better constrain its properties. The adopted observables are summarized in Table \ref{tab:HD156623}, including 
Str{\"o}mgren-Crawford ($uvby\beta$) photometry. 

The Str{\"o}mgren-Crawford photometric system is comprised of intermediate-band filters on either side of the Balmer break and two narrow-band filters that measure the strength of the H$\beta$ line \citep{Crawford58}. For early-type stars, the strength of the Balmer break, the continuum slopes in this region, and the strength of the H$\beta$ line are indicators of temperature and gravity. More specifically, for early A-type stars such as HD 156623, the temperature is sensitive to the index $a_0$:
\begin{equation}
\label{eqn1}
a_0 = 1.36\,(b - y)_0 + 0.36\,m_0 + 0.18\,c_0 - 0.2448
\end{equation}
while the surface gravity is highly sensitive to the reddening-free index 
$r_*$: 
\begin{equation}
\label{eqn2}
r_* = 0.35\,c_1 - 0.07\,(b - y) - (\beta - 2.565)
\end{equation}
where $m_0 = (v-b)_0 - (b-y)_0$ and $c_0 = (u-v)_0 - (v-b)_0$ \citep{Napiwotzki93}.

For dereddening the photometry and estimating the extinction due to 
interstellar dust, we adopt the reddening based on the regularly updated tri-dimensional reddening maps from the STILISM program \citep{Capitanio17, Lallement18}\footnote{\url{https://stilism.obspm.fr/}}: $E(B-V)$ = 0.008\,$\pm$\,0.018. 
Examining the results of \citet{McCall04} for low reddening
and dwarf stars in the color range -0.32 $<$ $(B-V)_o$ $<$ 1.5, one finds that the trend $A_V$/E($B-V$) $\simeq$ $R_V^{00}$ + 0.167\,$(B-V)_o$ is a reasonable approximation (where $R_V^{00}$ $\simeq$ 3.07 is quoted as the most
probably value for diffuse Galactic interstellar medium). Hence,
we adopt the ratio of total to selective extinction  $A_V$/E($B-V$) for our A1V
star to be 3.08, and estimate the extinction to be 
$A_V$ = 0.025\,$\pm$\,0.055. Using the E($B-V$) value, we use the 
``observed'' $uvby\beta$ extinction ratios reported in Table 2 of \citet{Fitzpatrick1999} to calculate the color excesses among the
Str{\"o}mgren-Crawford colors (see Table \ref{tab:uvby}). The color
excesses are all negligible, within 2$\sigma$ of zero. 

\begin{deluxetable}{ccc}
\tablecaption{Derived Photometric Parameters for HD 156623\label{tab:uvby}}
\tablehead{
\colhead{Parameter} & \colhead{Value (mag)} & \colhead{Reference}}
\colnumbers
\startdata
$A_V$    &  0.025\,$\pm$\,0.055 & 1\\ 
E($b-y$) &  0.0131\,$\pm$\,0.0092 & 2\\ 
E($m_1$) & -0.0042\,$\pm$\,0.0029 & 2\\ 
E($c_1$) &  0.0026\,$\pm$\,0.0018 & 2\\ 
E($u-b$) &  0.020\,$\pm$\,0.014 & 2\\ 
($b-y$)$_0$ & 0.0269\,$\pm$ 0.0096& 2\\ 
$m_0$ & 0.1972 $\pm$ 0.0050 & 2\\ 
$c_0$ & 0.9574 $\pm$ 0.0027 & 2\\  
$a_0$ & 0.035 $\pm$ 0.013 & 2\\
$r^*$ & -0.0058 $\pm$ 0.0012 & 2\\
\enddata
\tablecomments{
(1) This paper, calculated using STILISM E($B-V$) value and $R_V$ from \citet{McCall04},
(2) This paper, calculated using \citet{Fitzpatrick1999}.
}
\vspace{-0.4in}
\end{deluxetable}

\begin{figure}
\centering
\includegraphics[scale = 0.6]{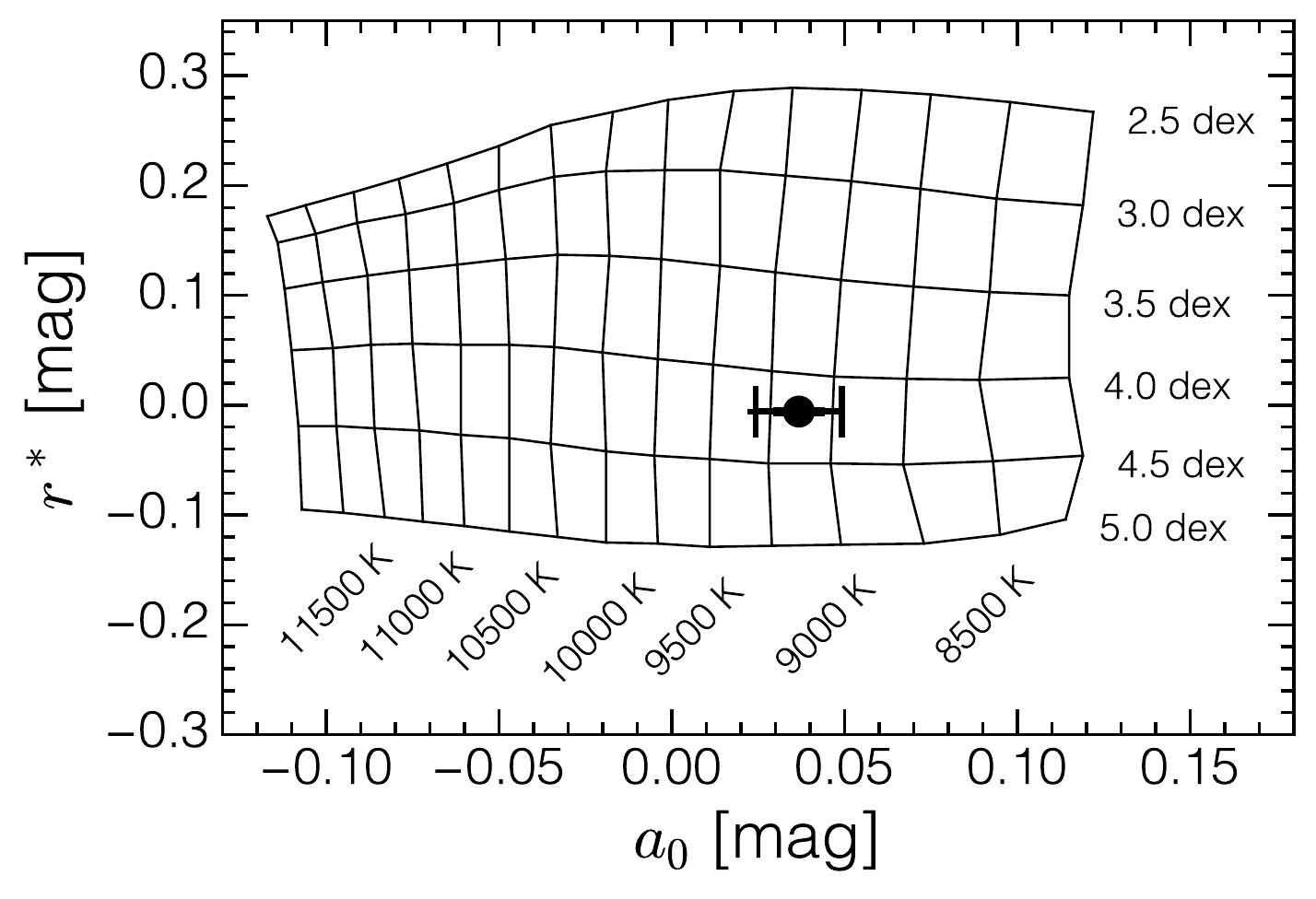}
\caption{Position of HD 156623 (black circle) in the $a_0-r^*$ plane with respect to synthetic atmosphere colors (grid lines) from \citet{CastelliKurucz2004,CastelliKurucz2006}.}
\label{fig:uvby}
\end{figure}

We then followed a procedure similar to that of \citet{DavidHillenbrand2015} and derived $T_\mathrm{eff}$ and $\log{g}$ through a two-dimensional linear interpolation in the $a_0-r^*$ plane using a synthetic color grid distributed by Fiorella Castelli\footnote{\url{http://wwwuser.oats.inaf.it/castelli/colors/uvbybeta/uvbybetap00k2odfnew.dat}}. The synthetic color grids are derived from ATLAS9 model atmospheres \citep{CastelliKurucz2004,CastelliKurucz2006}. We calculated uncertainties in the atmospheric parameters using Monte Carlo simulations, modeling the input photometry as normal distributions with widths equal to the errors reported in \citet{Paunzen15} and using the previously estimated $A_V$ extinction value. 
The dereddened $uvby\beta$ photometry and indices are presented in Table~\ref{tab:uvby} and the position of HD 156623 in the $a_0-r^*$ plane with respect to the models is shown in Figure~\ref{fig:uvby}.

From this analysis we estimate $T_\mathrm{eff} = 9040^{+240}_{-160}$~K and $\log{g} = 4.203^{+0.018}_{-0.011}$~dex. The values quoted here are the medians from the distributions resulting from the Monte Carlo simulations, while the errors quoted here are determined from the 16th and 84th percentiles of those distributions. We note that these errors are statistical in nature, and we have not attempted to account for any systematic uncertainties intrinsic to the synthetic color grids. We chose to instead adopt a systematic uncertainty of 0.14 dex for $\log{g}$ based on the analysis from \citet{DavidHillenbrand2015}. We adopt $T_\mathrm{eff} = 9040^{+240}_{-160}$~K and $\log{g} = 4.20\,\pm\,0.14$~dex.

\begin{deluxetable}{ccc}
\tablecaption{Parameters Derived for HD 156623\label{tab:HD156623derived}}
\tablehead{
\colhead{Parameter} & \colhead{Value} & \colhead{Units}}
\colnumbers
\startdata
T$_{\rm eff}$  &   9040$^{+240}_{-160}$   & K\\
$\log{g}$	   &   4.20\,$\pm$\,0.14      & dex\\
BC$_V$         &   -0.09\,$\pm$\,0.05     & mag\\
M$_V$          &   1.99\,$\pm$\,0.06      & mag\\
M$_{\rm bol}$  &   1.90\,$\pm$\,0.08      & mag\\
log(L/L$_{\odot})$ & 1.137\,$\pm$\,0.031  & dex\\
Radius         &   1.51\,$\pm$\,0.09      & R$_{\odot}$\\
Age            &   16\,$\pm$\,7           & Myr\\
$v_r$		& 3.8\,$\pm$\,6.9 & km/s\\
$v$sin$i$ & 88\,$\pm$\,2 & km/s\\
\enddata
\end{deluxetable}

We calculated the luminosity and radius of HD 156623 using the photometrically-derived parameters and Gaia parallax. We estimate the $V$-band bolometric correction using our estimated $T_{\rm eff}$ combined with various published $BC_V$ tables, all scaled to the new IAU 2015 bolometric magnitude system\footnote{\url{https://www.iau.org/static/resolutions/IAU2015$\_$English.pdf}, see also \citet{Mamajek15}. The IAU 2015 bolometric magnitude system is set so that $M_{bol}$ = 4.74 mag corresponds to the nominal solar luminosity of 3.828\,$\times$\,10$^{26}$ W.}. Here
we adopt the solar apparent $V$ magnitude from \citet{Torres10} (-26.76\,$\pm$\,0.03 mag), which when combined with the IAU 2015 apparent bolometric magnitude (-26.832 mag), equates to a solar $BC_V$ = -0.072\,$\pm$\,0.03 mag.
Comparing the $BC_V$ scales from several studies \citep{Balona94,Flower96,Bessell98,Bertone04}, adjusting the scales slightly to reproduce
$BC_V$(5772\,K) = -0.072 mag, and accounting for the $T_{\rm eff}$ uncertainties, we estimate the $V$-band bolometric correction for HD 156623 to be -0.09\,$\pm$\,0.05 mag. Accounting for the extinction ($A_V$ = 0.025\,$\pm$\,0.055 mag),
we derive bolometric flux $f_{bol}$ = 35.11\,$\pm$\,2.42 pW\,m$^{-2}$, 
apparent bolometric magnitude $m_{bol}$ = 7.139\,$\pm$\,0.075 mag, 
absolute bolometric magnitude $M_{bol}$ = 1.898\,$\pm$\,0.077 mag (IAU 2015 system), and absolute magnitude $M_V$ = 1.988\,$\pm$\,0.058 mag. 
We derive bolometric luminosity log(L/L$_{\odot}$) = 1.137\,$\pm$\,0.031 dex 
or 13.71$^{+1.01}_{-0.94}$ $L_{\odot}$, on the IAU 2015 scale where $L_{\odot}$ = 3.828\,$\times$\,10$^{26}$ W. 
Combined with the previously estimated effective temperature (T$_{\rm eff}$ = 9040$^{+240}_{-160}$~K),
we estimate the stellar radius to be 1.51\,$\pm$\,0.09 $R_{\odot}$, where $R_{\odot}$
is the IAU 2015 nominal solar radius of 695,700\,km. These estimated stellar parameters are summarized in Table \ref{tab:HD156623derived}.

To provide a consistency check on the parameters derived for HD 156623, we used the \texttt{isoclassify} code from \citet{Huber2017}, which uses the MIST evolutionary models \citep{Choi2016, Dotter2016} as well as the three-dimensional reddening map of \citet{Green2015} implemented in the \texttt{mwdust} package \citep{Bovy2016}. The \texttt{isoclassify} code has two modes: direct and grid. In the direct mode, the code takes an input parallax, $K$-band magnitude and priors on $T_\mathrm{eff}$, $\log{g}$, and [Fe/H] in order to determine the luminosity and radius using theoretical $K$-band bolometric corrections. The parallax is used to generate a distance posterior distribution, which is then used to calculate reddening given the star's position and a 3D dust map. In the grid mode, the code does not rely on a reddening map but instead treats extinction as a free parameter. Using a fine grid of MIST isochrones, the code calculates reddened photometry in the 2MASS $JHK$, Tycho $B_TV_T$, and Sloan $griz$ passbands for given values of $A_V$ by interpolating the \citet{Cardelli1989} extinction law. The code then integrates over all isochrone points to find the maximum likelihood model that matches various combinations of input observables (e.g. colors). 

The $\log{g}$ derived from $uvby\beta$ photometry is consistent within 1$\sigma$ of the value found using the \texttt{isoclassify} grid mode (4.28$\,\pm\,$0.01 dex). The radii (R$_{\rm grid}$ = 1.635$\,\pm\,$0.026 R$_{\odot}$, R$_{\rm direct}$ 1.618$\,\pm\,$0.092 R$_{\odot})$, $T_{\rm eff}$ (T$_{\rm eff,grid}$ = 8920$\,\pm\,$230 K, T$_{\rm eff,direct}$ = 9040$\,\pm\,$240 K), and luminosity (log(L/L$_{\odot,\rm grid}$) = 1.18$\,\pm\,$0.04 dex, log(L/L$_{\odot,\rm direct}$) = 1.20$\,\pm\,$0.01 dex) values found using both modes are consistent with one another and the results in Table \ref{tab:HD156623derived} within $\sim$2$\sigma$.

\subsection{Spectral Analysis}
The FEROS spectrum of HD 156623 was downloaded in the FITS file format and analyzed with the SPLAT-VO tool\footnote{\url{http://star-www.dur.ac.uk/$\sim$pdraper/splat/splat-vo/}}. The wavelength range of 3900 -- 6800 \AA\, was used to avoid the Balmer jump and near-IR telluric lines interfering with the normalization process. Within SPLAT-VO, the spectrum was normalized by carefully using the line draw tool to interpolate an estimated continuum. The normalized spectrum was then analyzed to determine several properties of the star.
\begin{deluxetable}{ccc}
\tablecaption{Radial Velocity Spectral Lines\label{tabspec}}
\tablehead{
\colhead{Element} & \colhead{Air Wavelength (\AA)} &  \colhead{Measured Wavelength (\AA)}}
\colnumbers
\startdata
H$\alpha$  &   6562.800 & 6562.779\,$\pm$\,0.034\\
H$\beta$	   &   4861.330 & 4861.387\,$\pm$\,0.033\\
H$\gamma$         &   4340.470 & 4340.303\,$\pm$\,0.030\\
H$\delta$          &   4101.760 & 4101.946\,$\pm$\,0.025\\
\ion{Ca}{2} K  &  3933.636 & 3933.654\,$\pm$\, 0.050\\
\ion{Si}{2}& 4128.070& 4128.053\,$\pm$\,0.045\\
\ion{Sr}{2}&   4215.519& 4215.577\,$\pm$\,0.015\\
\ion{Ca}{1}&   4226.728& 4226.882\,$\pm$\,0.011\\
\ion{Fe}{2}& 4233.167& 4233.273\,$\pm$\,0.017\\
\ion{Sc}{2} & 4246.820& 4246.921\,$\pm$\,0.013\\
\ion{Fe}{2}& 4250.429&4250.509\,$\pm$\,0.011\\
\ion{Fe}{2}& 4271.400&4271.638\,$\pm$\,0.012\\
\ion{Fe}{2}& 4289.775&4289.962\,$\pm$\,0.011\\
\ion{Fe}{1}& 4404.750&4404.717\,$\pm$\,0.014\\
\ion{Fe}{2}& 4416.339&4416.319\,$\pm$\,0.053\\
\ion{Ti}{2} & 4468.507&4468.456\,$\pm$\,0.022\\
\enddata
\end{deluxetable}

The Balmer series, \ion{Ca}{2} K, and several metal lines in the spectrum (summarized in Table \ref{tabspec}) were fit in SPLAT-VO with Gaussian, Lorentzian, and Voigt profiles. The best profile for each individual line was determined by the lowest rms of their fit. The profiles returned central wavelengths for each line, which were used to estimate a radial velocity for the star. The central wavelengths were compared to air wavelengths available in the NIST Atomic Spectra database\footnote{\url{http://physics.nist.gov/PhysRefData/ASD/lines\_form.html}}. Based on this analysis, the radial velocity for HD 156623 was determined to be $v_{r}$ = 3.8\,$\pm$\,6.9 km/s, which was reported in Table \ref{tab:HD156623derived}. The best value for the radial velocity was determined from the average value of the $v_{r}$ measured from each line; the uncertainty was propagated from the fit errors on the measured central wavelengths. Notably absent from this analysis were the H$\epsilon$ and \ion{Ca}{2} H lines and the \ion{Mg}{2} doublet at 4481 \AA\,. Both sets of lines were blended in this spectrum and led to radial velocity values that were incompatible with the values measured in the other lines. This led to their exclusion from this analysis.

To estimate $v$sin$i$ for HD 156623, several synthetic spectra from the POLLUX\footnote{\url{http://pollux.graal.univ-montp2.fr/}} service were generated. The POLLUX \citep{Palacios10} service provides synthetic ATLAS12 \citep{Kurucz05} atmospheres for early-A stars with T$_{\rm eff}$ in steps of 100 K, $\log{g}$ in steps of 0.1 dex, and [Fe/H] in steps of 0.5 dex. Within the POLLUX service, the ATLAS12 models can be convolved with a synthetic broadening profile based on the work by \citet{Gray05}. This broadening profile is determined by an input macroturbulent velocity, rotational velocity, instrument profile, and radial velocity. To extract a $v$sin$i$, we chose to fit these artificially broadened ATLAS12 models from the POLLUX service to the well-normalized H$\beta$ and narrow metal lines present in the spectrum. This method for estimating $v$sin$i$ is similar to Method 2 described in \citet{Brown97S} and is applicable for 50 km/s $\le$ $v$sin$i$ $\le$ 200 km/s.

\begin{figure}
\centering
\includegraphics[scale=0.6]{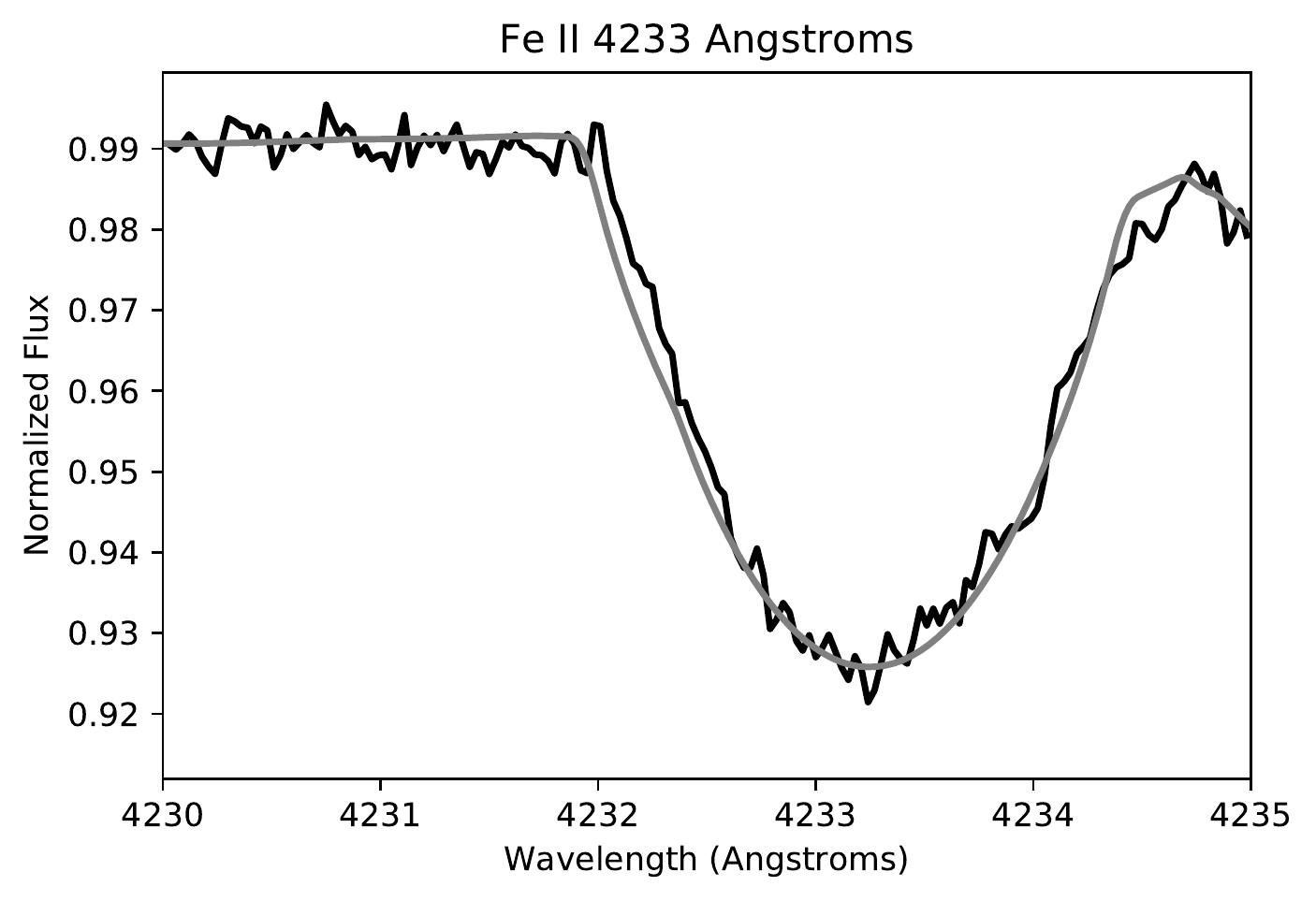}
\caption{An example fit using the \ion{Fe}{2} line at 4233 \AA\,. The black line represents the normalized FEROS spectrum. The gray line is the best fit artificially broadened ATLAS12 model spectrum (T$_{\rm eff}$ = 9200 K, $\log{g}$ = 4.2 dex, [Fe/H] = -0.5, and $v$sin$i$ = 87 km/s).}
\label{fig:specline}
\end{figure}

For a simple estimate of $v$sin$i$, we assumed a broadening profile due to rotation only, i.e., $v_{macro}$ = 0 km/s and negligible instrument profile. The lines used to determine $v$sin$i$ were selected based on their strength in the HD 156623 spectrum and their use in previous studies \citep[H$\beta$, \ion{Fe}{2} 4233 \AA\,, \ion{Ca}{1} 4226 \AA\,, \ion{Mg}{1} 4702 \AA:][]{Ramella89,Royer02,Royer14,Zwintz14}. The first line analyzed was the \ion{Fe}{2} \AA\, line. This line provided a strong means of measuring a close value for the atmospheric parameters and broadening profile simultaneously. With initial guesses based on the Str{\"o}mgren-derived parameters, we varied T$_{\rm eff}$, $\log{g}$, [Fe/H], and $v$sin$i$ in steps of 100 K, 0.1 dex, 0.5 dex, and 10 km/s respectively. Each fit was evaluated with a $\chi^2$ value. This located a starting atmosphere of T$_{\rm eff}$ = 9200 K, $\log{g}$ = 4.2 dex, [Fe/H] = -0.5 dex, and $v$sin$i$ = 90 km/s.

This initial guess of $v$sin$i$ = 90 km/s was then varied in steps of $\pm$\,2 km/s and then $\pm$ 1 km/s. This resulted in a best fit $v$sin$i$ for the \ion{Fe}{2} line of T$_{\rm eff}$ = 9200 K, $\log{g}$ 4.2 dex, [Fe/H] = -0.5 dex, and $v$sin$i$ = 87 km/s. The \ion{Fe}{2} 4233 \AA\, line and its best fit were plotted in Figure \ref{fig:specline} as an example. The black line in Figure \ref{fig:specline} represents the measured FEROS spectrum and the gray line represents the best fit theoretical spectrum. The other two metal lines were best fit with this same atmosphere and the H$\beta$ line was better fit with T$_{\rm eff}$ = 9300 K and $v$sin$i$ = 90 km/s. Using these four lines, we adopt $v$sin$i$ = 88 $\pm$ 2 km/s. Our velocity results are consistent with \citet{Rebollido18}.
\section{Results \& Discussion \label{sec:discussion}}

\subsection{Group Membership and Age \label{stellarage}}

\begin{figure}
\centering
\includegraphics[scale=0.4]{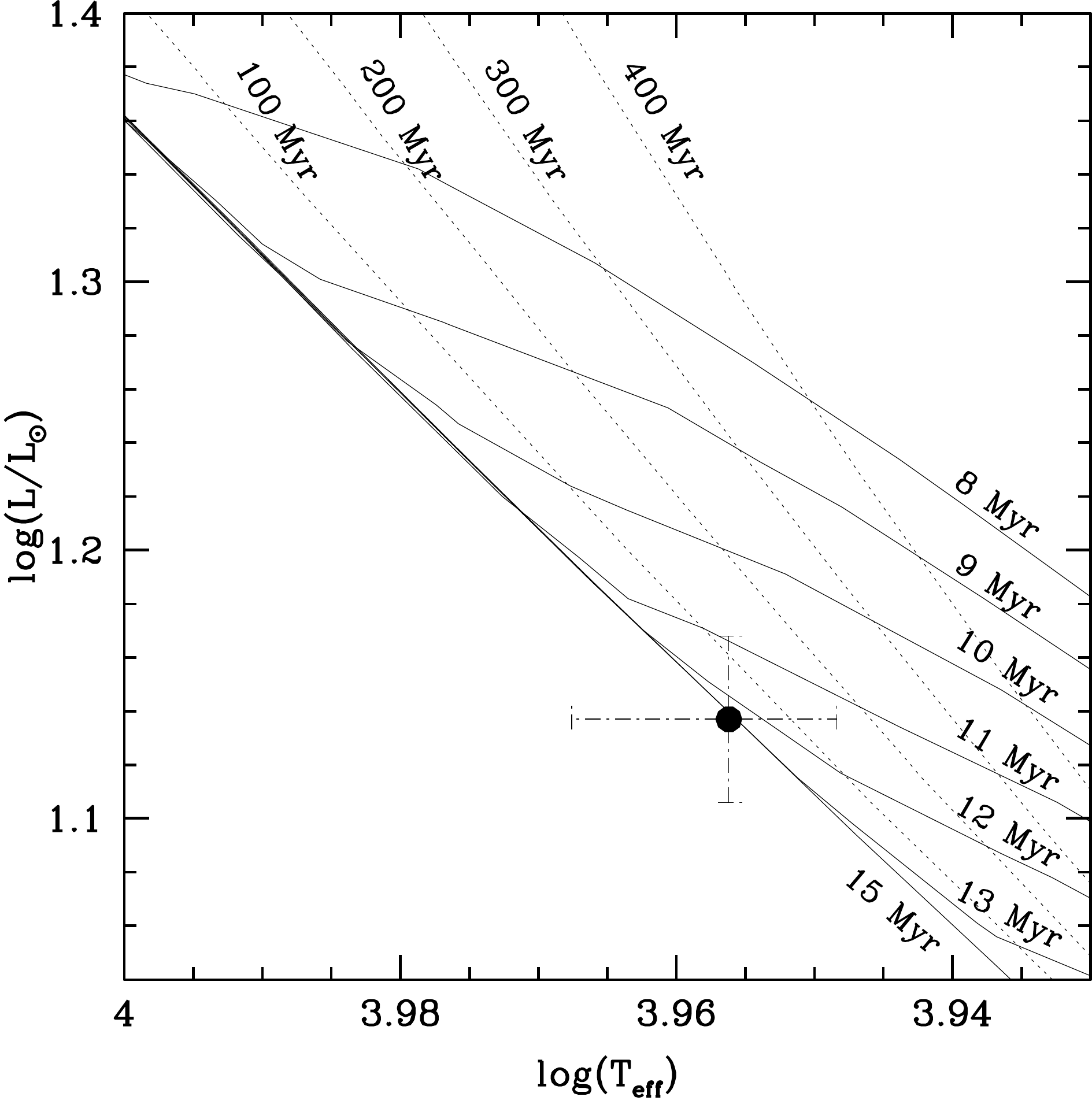}
\caption{The HR diagram position of HD 156623 plotted along with PARSEC track solar composition isochrones. The position on the HR diagram is consistent with a ZAMS star between $\sim$11 Myr and $\sim$200 Myr.}  
\label{fig:156HR}
\end{figure}

The results and values in Table \ref{tab:HD156623derived} were used to evaluate the age of HD 156623 based on its HR diagram position and perform a Sco-Cen membership analysis. The HR diagram position of the star is plotted {(in Figure \ref{fig:156HR})} along with solar composition isochrones from the PARSEC tracks \citep{Marigo17}\footnote{Isochrones generated via the online CMD 3.0 tool at \url{http://stev.oapd.inaf.it/cgi-bin/cmd}, using initial composition X = 0.7092, Y = 0.2755, Z = 0.0152.}. The HRD position for the star is consistent with a ZAMS star, likely older than $\sim$11 Myr and younger than $\sim$200 Myr.

Applying the BANYAN $\Sigma$ kinematic membership tool\footnote{\url{http://www.exoplanetes.umontreal.ca/banyan/banyansigma.php}} (which assesses membership probabilities for stars based on position and kinematic data \citep{Gagne18}) with the Gaia DR2 astrometry, we found a membership probability of 88.6\%\, for the HD 156623 to belong to the Upper Centaurus-Lupus (UCL) subgroup of Sco-Cen \citep{deZeeuw99,Mamajek02,Preibisch08,Pecaut16}. BANYAN $\Sigma$ also estimated a 11.4\%\, probability that the star is a field star, and a probability of $<$0.1\%\, for it to belong to any of the other known kinematic groups within 150\,pc included in the code's database. If the star belongs to UCL, its predicted radial velocity is 3.2\,km/s, which is consistent with our measured value of 3.8\,$\pm$\,6.9 km/s. UCL has a mean age of 16\,$\pm$\,2\,Myr, however the group covers tens 
of parsecs, and appears to have an intrinsic age spread of approximately $\pm$7 Myr \citep{Pecaut16}. UCL surrounds the Lupus clouds,
with $\sim$5-10 Myr-old stars in the vicinity (but outside of the star-forming clouds), and pockets of UCL appear to have ages as old $\sim$20-25 Myr. We surmise that HD 156623 is most likely an outlying member of UCL, with age $\sim$16\,$\pm$\,7 Myr, based on the intrinsic scatter in the ages of UCL members from \citep{Pecaut16}, and consistent with the HR diagram position. This suggests that the star is ZAMS, or possibly slightly pre-MS if it is younger than about $<$12 Myr.

\subsection{$\delta$ Scuti Frequencies \label{dsfreq}}

\subsubsection{Reported Frequencies \label{reporting}}

The analysis of {\S \ref{extracting}} recovered 16 p-mode frequencies for HD 156623. Since the noise characteristics of bRing were not completely understood {(see \S \ref{noise})}, this study was conservative when reporting frequencies as astrophysical. Due to the design of bRing, we had the ability to check for each frequency in each camera individually when selecting frequencies. These 16 frequencies were subsequently broken into a group of 9 confirmed frequencies {(Table \ref{tab:dHD156623})} and 7 candidate frequencies ({Table \ref{tab:2dHD156623}}). The 9 confirmed frequencies were detected in data taken by at least one camera at each site. This criterion guaranteed that these frequencies are not single-camera and single-site systematics being mistakenly reported as real astrophysical frequencies intrinsic to the star. The 7 candidate frequencies were only detected in the South African West Camera (SAW), which had observed the star the most with $\sim$8500 data points. These frequencies reported in {Table \ref{tab:2dHD156623}} could be either real astrophysical frequencies or unknown systematics intrinsic to either SAW or the South African bRing station itself. We also included in {Table \ref{tab:dHD156623}} the number of cameras each frequency was detected in. The cameras that missed some of the frequencies were either the Australia West or South Africa East cameras, which both observed the star the least at their respective site.

\begin{deluxetable}{lllc}
\tablecaption{9 Confirmed $\delta$ Scuti Frequencies\label{tab:dHD156623}}
\tablehead{
\colhead{Frequency}& \colhead{Amplitude} & \colhead{Phase} & \colhead{\# of cameras}\\ \colhead{($d^{-1}$)} & \colhead{(mmag)} & \colhead{(Rad)} & \colhead{--}}
\startdata
71.143\,$\pm$\,0.002 & 6.63\,$\pm$\,1.23 & 0.565\,$\pm$\,0.001 & 4\\
67.005\,$\pm$\,0.002 & 2.60\,$\pm$\,0.99 & 0.542\,$\pm$\,0.001 & 4\\
67.306\,$\pm$\,0.002 & 2.31\,$\pm$\,0.65 & 0.114\,$\pm$\,0.001 & 4\\
63.562\,$\pm$\,0.002 & 2.18\,$\pm$\,1.29 & 0.570\,$\pm$\,0.001 & 3\\
63.426\,$\pm$\,0.002 & 1.70\,$\pm$\,0.34 & 0.072\,$\pm$\,0.001 & 4\\
59.002\,$\pm$\,0.003 & 1.69\,$\pm$\,0.33 & 0.413\,$\pm$\,0.001 & 3\\
63.701\,$\pm$\,0.003 & 1.58\,$\pm$\,0.31 & 0.637\,$\pm$\,0.001 & 3\\
59.970\,$\pm$\,0.003 & 1.03\,$\pm$\,0.09 & 0.556\,$\pm$\,0.001 & 2\\
70.880\,$\pm$\,0.003 & 1.00\,$\pm$\,0.18 & 0.258\,$\pm$\,0.001 & 2\\
\enddata
\tablecomments{Mean Epoch 2458193.3 HJD}
\end{deluxetable}

\begin{deluxetable}{lll}
\tablecaption{7 Candidate $\delta$ Scuti Frequencies\label{tab:2dHD156623}}
\tablehead{
\colhead{Frequency}& \colhead{Amplitude} & \colhead{Phase}\\ \colhead{($d^{-1}$)} & \colhead{(mmag)} & \colhead{(Rad)}}
\startdata
93.261\,$\pm$\,0.002 & 1.04\,$\pm$\,0.78 & 0.896\,$\pm$\,0.001\\
60.700\,$\pm$\,0.002 & 0.95\,$\pm$\,0.78 & 0.873\,$\pm$\,0.001\\
75.211\,$\pm$\,0.002 & 0.82\,$\pm$\,0.78 & 0.882\,$\pm$\,0.002\\
66.414\,$\pm$\,0.002 & 0.79\,$\pm$\,0.78 & 0.457\,$\pm$\,0.002\\
64.179\,$\pm$\,0.002 & 0.75\,$\pm$\,0.78 & 0.581\,$\pm$\,0.002\\
89.250\,$\pm$\,0.002 & 0.73\,$\pm$\,0.78 & 0.434\,$\pm$\,0.002\\
56.116\,$\pm$\,0.002 & 0.70\,$\pm$\,0.78 & 0.198\,$\pm$\,0.002\\
\enddata
\tablecomments{Mean Epoch 2458193.3 HJD}
\end{deluxetable}

The frequencies and phases reported in {Table \ref{tab:dHD156623}} were the frequencies and phases detected using {\tt Period04} on the composite periodograms. The reported amplitudes in {Table \ref{tab:dHD156623}} were from averaging the composite and individual camera detections uncovered by {\tt Period04}. The uncertainties for the frequencies and amplitudes were calculated based on the measurements made in each camera and the composite light curve. This resulted in uncertainties that were 1 order of magnitude larger in frequency than uncertainties calculated by the methods of \citet{Montgomery99}, but the presence of remaining unidentified systematics due to the high observing cadence and stationary nature of bRing requires a more conservative estimate of the noise until these systematics are investigated further and removed. The phase uncertainties, however, were calculated via \citet{Montgomery99} using the amplitudes from {Table \ref{tab:dHD156623}}. The frequencies, amplitudes, and phases in {Table \ref{tab:2dHD156623}} were reported as detected in SAW. The uncertainties for the frequency and amplitude were computed using the uncertainties from {Table \ref{tab:dHD156623}}. The uncertainty in the phase for {Table \ref{tab:2dHD156623}} was similarly computed using \citet{Montgomery99}. The amplitude uncertainties were rather large compared to the reported amplitudes. The culprit was likely slight temporal systematics between the cameras.

To ensure that the frequencies were exclusive to HD 156623, the 3 nearest neighbors in the bRing catalog (HD 157661 $\sim$68 pixels apart, HD 156274 $\sim$75 pixels apart, HD 156293 $\sim$80 pixels apart) were used as check stars. They were put through the same data reduction and periodogram analysis discussed in \S \ref{sec:data}. These stars were chosen as they should be subject to similar sources of global and local noise on the CCD due to their proximity. Therefore, any unique signals detected in the periodogram of HD 156623 should be independent of any signals detected in the periodograms of these stars. We found no evidence of pulsations within the 50 $d^{-1}$ -- 100 $d^{-1}$ frequency range of these three check stars.

\begin{figure}
\centering
\includegraphics[scale=0.2]{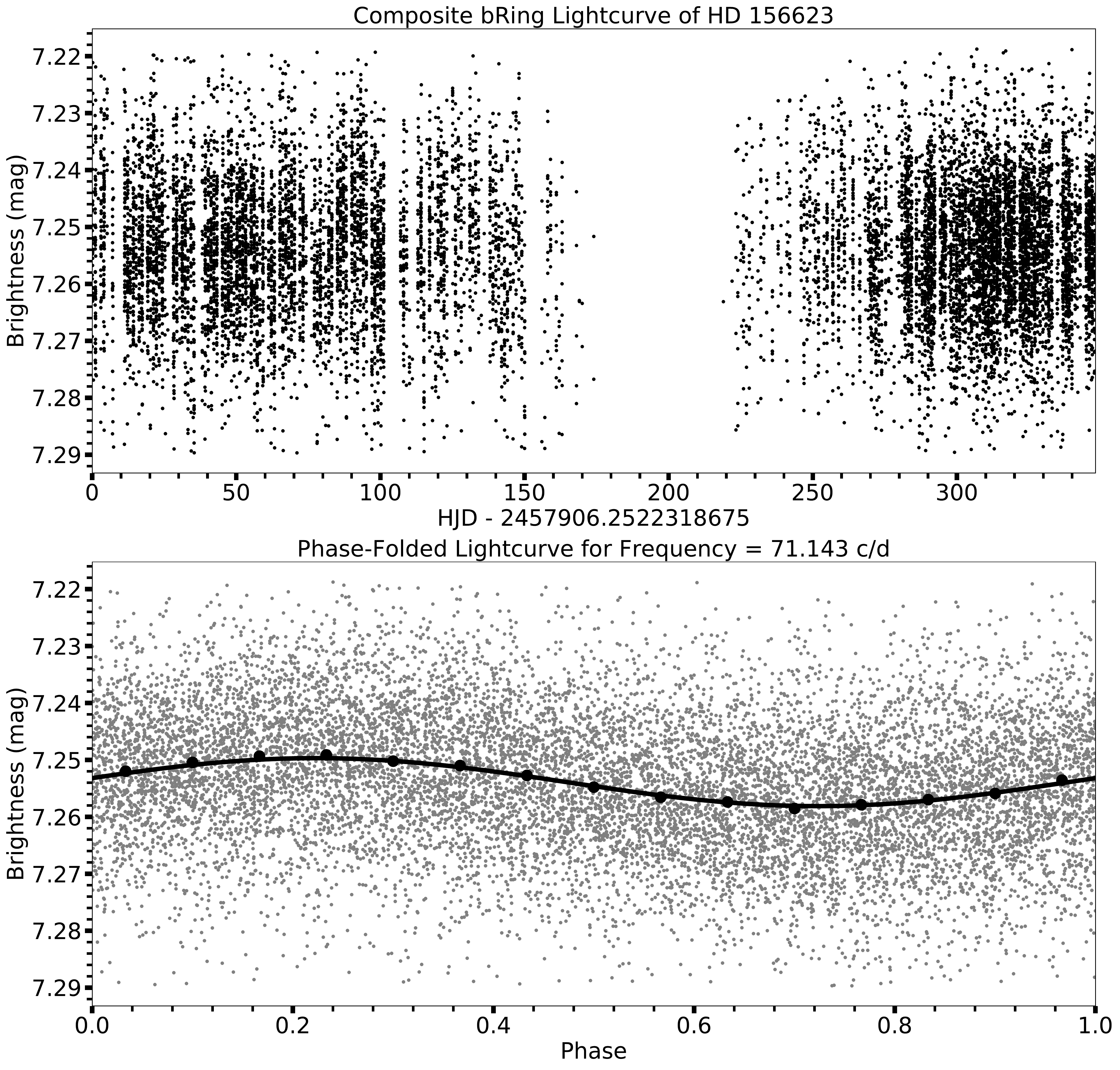}
\caption{The top panel contains the composite bRing light curve of HD 156623. The bottom panel contains the composite bRing light curve (gray dots) phase-folded onto the primary 71.143 $d^{-1}$ frequency. The black circles are 15 median binned data points and the black curve is the sinusoidal solution found by $\tt{Period04}$.}
\label{lc}
\end{figure}

In the top panel of Figure \ref{lc}, we plotted the full composite light curve of HD 156623. In the bottom panel of Figure \ref{lc}, we plotted a phase-folded light curve of the primary frequency (71.143 $d^{-1}$) in Figure \ref{lc}. This bottom plot includes the bRing data in gray, 15 evenly spaced median-binned data points represented by black circles, and the sinusoidal solution from $\tt{Period04}$ plotted as a black line.

\subsubsection{Interpreting the Frequencies}\label{fanalysis}
\begin{figure*}
\centering
\includegraphics[height = 2.0\columnwidth,width=2.0\columnwidth]{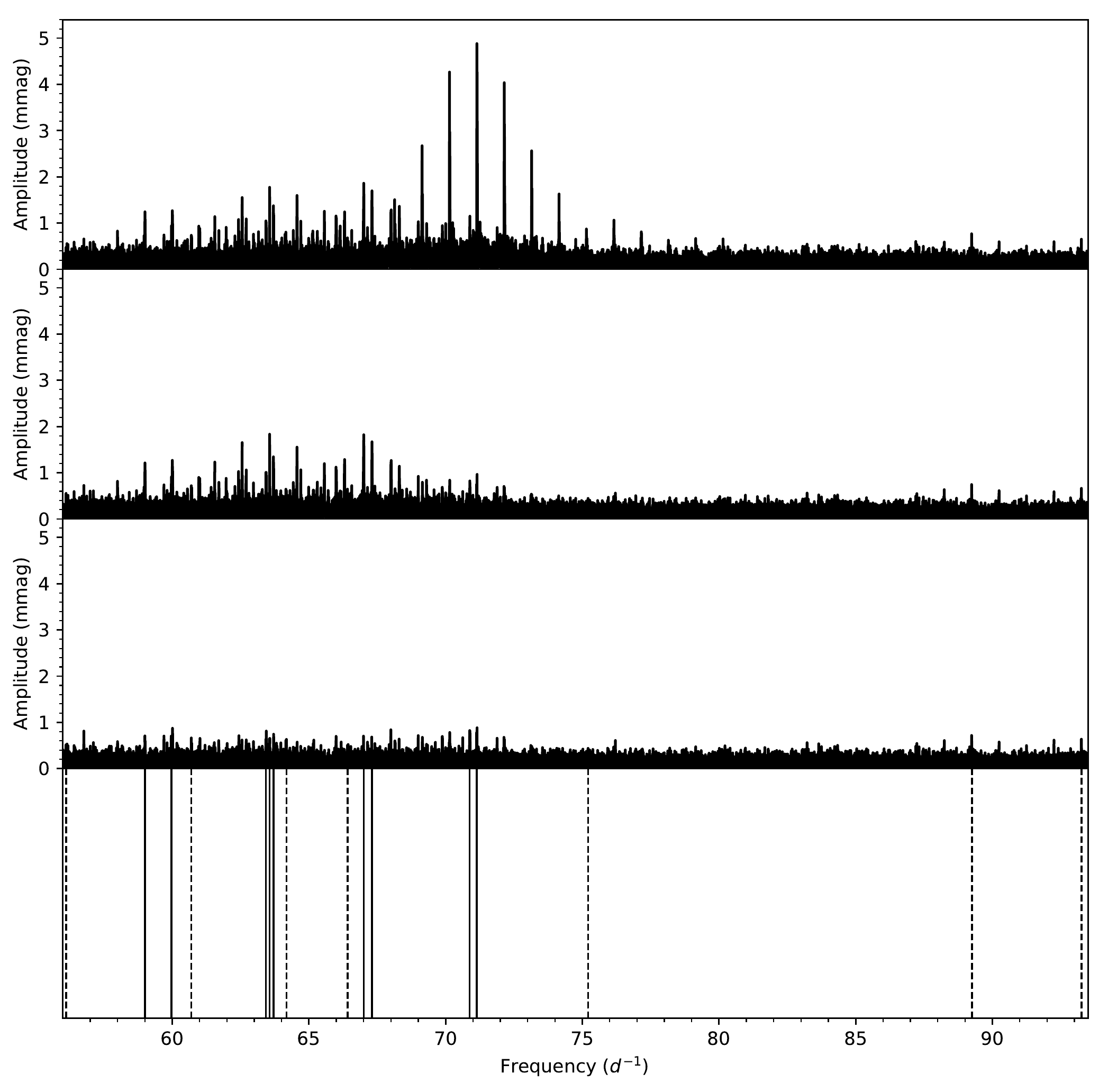}
\caption{The Lomb-Scargle periodogram of the HD 156623 composite light curve is plotted in the top panel. The second panel shows the residual periodogram after the removal of the primary frequency. The third panel shows the residual noise periodogram. In the bottom panel, the projections of the 9 confirmed frequencies from Table \ref{tab:dHD156623} were plotted with solid lines and the 7 candidate frequencies from Table \ref{tab:2dHD156623} were plotted with dashed lines. The frequency span is between 56 $d^{-1}$ and 93.5 $d^{-1}$ to focus on the $\delta$ Scuti features. It should be noted that the amplitudes plotted in the top panel periodogram correspond to the amplitudes from the periodogram of the composite light curve and not the reported amplitudes from Tables \ref{tab:dHD156623} and \ref{tab:2dHD156623}.}
\label{fig:HD156623}
\end{figure*}
HD 156623 has been a photometric standard star for decades. It is expected that any new detected pulsations in the star should be comparable or smaller than the best value of the $V$ magnitude and its associated uncertainty \citep[$V$ = 7.24 $\pm$ 0.007 mag;][]{Mermilliod97}. Considering the confirmed frequencies only, one finds the sum of the amplitudes to be $\sim$20 mmag, which is $\sim$3 times lower than the SEM of the $V$ magnitude hence it is unsurprising that the $\delta$ Scuti pulsations were previously missed.

In the top panel of {Figure \ref{fig:HD156623}}, the periodogram of the HD 156623 composite light curve was plotted between 56 $d^{-1}$ and 93.5 $d^{-1}$, within which the $\delta$ Scuti pulsations are contained. The second panel shows the periodogram after removal of the primary frequency and the third shows the residual periodogram after the removal of all frequencies above the noise threshold. The projections of the 9 reported frequencies from Table \ref{tab:dHD156623} were plotted with solid lines in the bottom panel of {Figure \ref{fig:HD156623}} along with the projections of the 7 candidate frequencies from {Table \ref{tab:2dHD156623}} plotted with dashed lines. The main frequency (71.143 $d^{-1}$) generated several daily aliases, which are obvious in the periodogram. There were also several frequencies that seemed real; upon closer inspection, these appeared to be aliases between the peaks (several strong beat frequencies were also generated and are outside the window of {Figure \ref{fig:HD156623}}). The reported frequencies are not aliases of each other or linear combinations of one another. They were also well-resolved, despite our large uncertainties. In HD 156623, we clearly observed high-order radial modes \citep[the fundamental radial should not exceed 25 ${d^{-1}}$;][]{Zwintz11}. 

\begin{figure}
\centering
\includegraphics[scale=0.6]{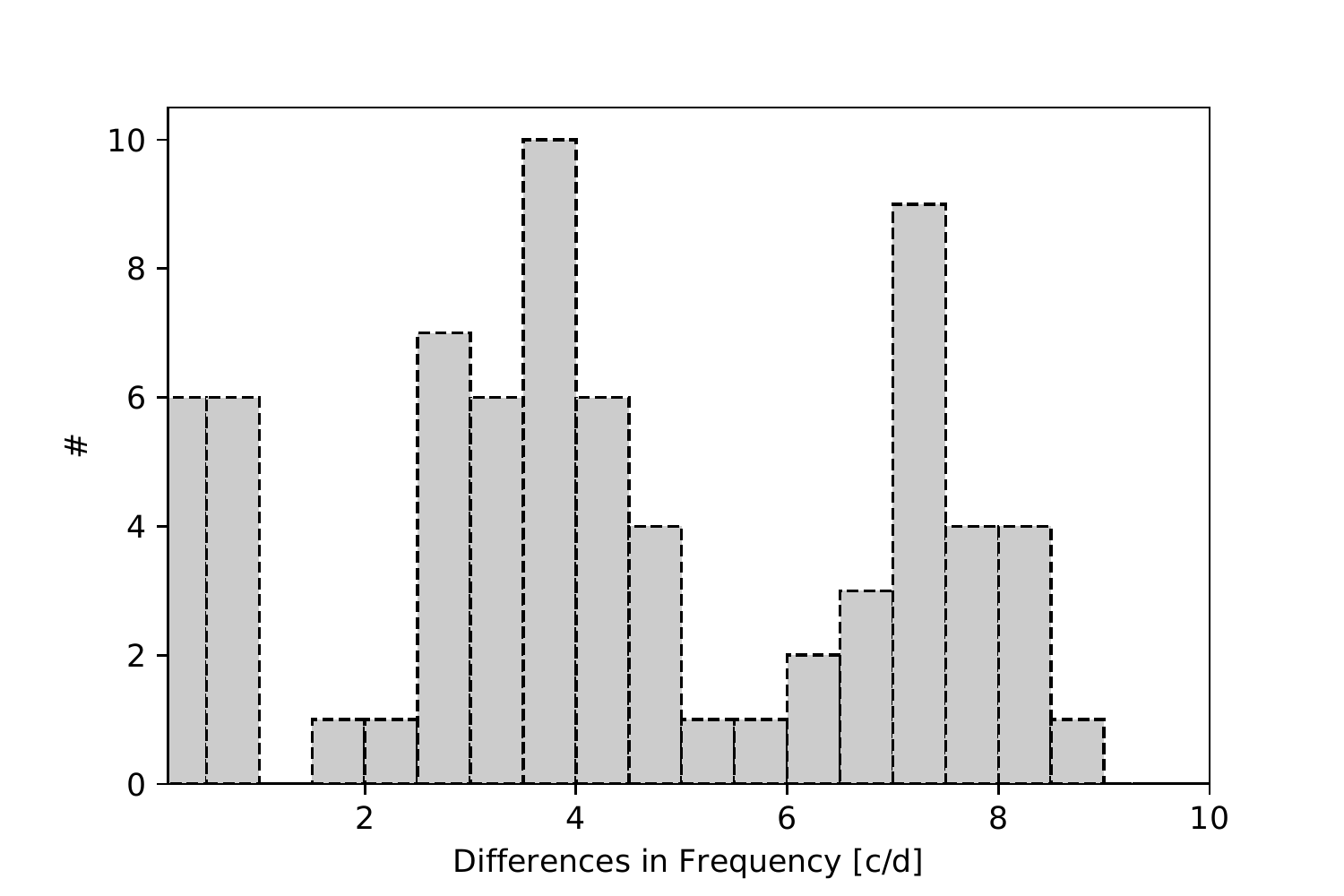}
\caption{A histogram of frequency differences that used all 16 frequencies detected by this study (0.5 $d^{-1}$ bins). This histogram shows strong evidence of regularity at the spacings of 3.75 $d^{-1}$, 7.25 $d^{-1}$, and 2.75 $d^{-1}$.}
\label{histo}
\end{figure}

There was evidence of grouping between all 16 frequencies \citep[frequencies clustered around each other;][]{Kurtz14}. Without a proper model and 7 of our frequencies only reaching the candidate status, we deduced a possible interpretation of how the frequencies may be grouped based on a visual inspection of Figure \ref{fig:HD156623}. The main frequency (71.143 $d^{-1}$) appeared to exist in a doublet with the frequency 70.880 $d^{-1}$. We also saw evidence for two sets of triplets (\{67.005 $d^{-1}$, 67.306 $d^{-1}$, 66.414 $d^{-1}$\} and \{59.002 $d^{-1}$, 59.970 $d^{-1}$, 60.700 $d^{-1}$\}) and 1 quadruplet \{63.562 $d^{-1}$, 63.426 $d^{-1}$, 63.701 $d^{-1}$, 64.179 $d^{-1}$\}. The rest of the frequencies appeared to be singlets. The frequencies reported by this analysis showed strong evidence of regularity \citep[frequencies with common separation;][]{Zwintz11}. We plotted a histogram of the differences between all 16 frequencies -- using bins of size 0.5 $d^{-1}$ -- in {Figure \ref{histo}}; we found evidence of regularity for three different separations: 3.75 $d^{-1}$, 7.25 $d^{-1}$, and 2.75 $d^{-1}$. To generate an accurate interpretation of these regularities and groupings, one would need to develop a model similar to the one from \citep{Zwintz14}, which is beyond the scope of this work.

To check for temporal variations in the frequencies, amplitudes, and phases, the data were broken up into $\sim$2 month segments in SAW (the best camera to individually sample the star with $\sim$8500 points). All three parameters for each set of frequencies stayed consistent (within uncertainty) throughout each segment. Therefore, period, amplitude, and phase variations were not detected in this study on the timescale of $\sim$1 year.

\subsubsection{Evolutionary Stage Analysis \label{evolstage}}

A critical parameter in the seismic analysis of pre-MS and ZAMS stars is the acoustic-cutoff frequency \citep{Zwintz14c}. This corresponds to the highest frequency ($f_{max}$) pressure mode in the star \citep{Zwintz14}. \citet{Casey11} predicts that the highest acoustic-cutoff frequencies occur as the star approaches ZAMS. It is also expected that the acoustic-cutoff frequency scales with the frequency of maximum power \citep{Aerts10}. The echographic study of \citet{Zwintz14c} used pre-MS and ZAMS $\delta$ Scutis to show that the hottest and most evolved stars are consistent with these predictions. For HD 156623, we found that the acoustic-cutoff frequency is also the frequency of maximum power. This is similar to HD 34282, although $f_{max}$ for HD 34282 is higher \citep[79.423 $d^{-1}$;][]{Casey13}. This implied -- like HD 34282 -- that HD 156623 should be consistent with the models and implications predicted by \citet{Casey11} and \citet{Aerts10} and be consistent with the results of \citet{Zwintz14c} if the age and evolutionary status estimations from \S \ref{stellarage} were correct.

\begin{figure*}
\includegraphics[angle=270,width=\textwidth,trim =0cm 0cm 1.7cm 0.7cm,clip]{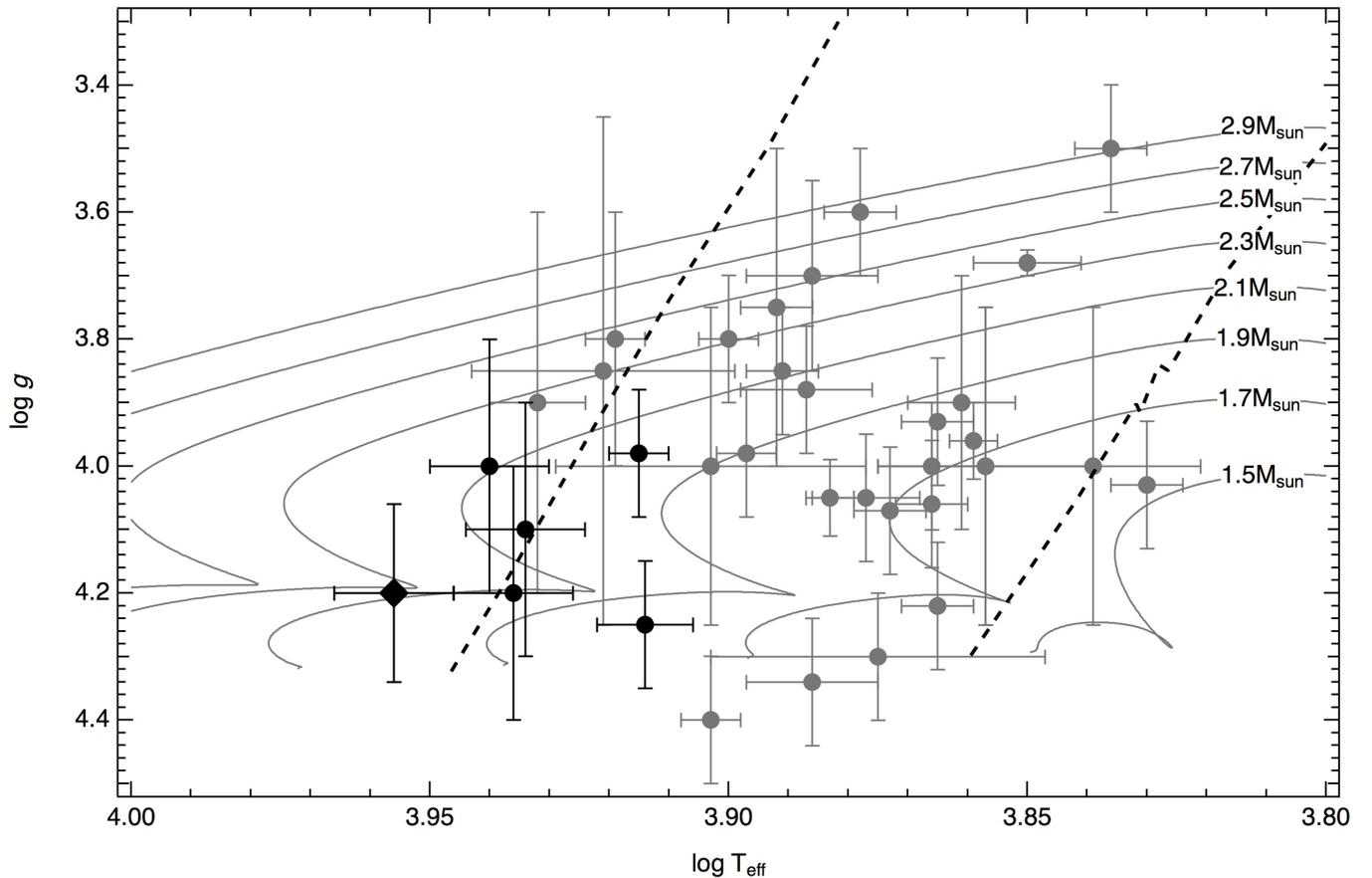}
\caption{A Kiel diagram of known pre-MS and early ZAMS $\delta$ Scuti stars (gray and black circles) including HD 156623 marked with a diamond symbol. The stars in black are the hottest stars with the highest excited p-mode frequencies known (including HD 156623 (A1\,V), $\beta$ Pic (A6\,V), HD 261711 (A2\,V), HD 144277 (A1\,V), HD 34282 (A0.5\,V) and HD 37357 (A1\,V)) as described by \citet{Zwintz14c}. The solid lines are the pre-MS evolutionary tracks from \citet{Guenther09} from 1.5 $M_{\odot}$ to 2.9 $M_{\odot}$ in steps of 0.2 $M_{\odot}$. The theoretical blue edge (left dashed line) and empirical red edge (right dashed line) of the classical $\delta$ Scuti instability strip were also plotted \citep{Breger98}.}
\label{kiel}
\end{figure*}

To check this, we used the estimated $T_\mathrm{eff}$ and $\log{g}$ from {Table \ref{tab:HD156623derived}} and the acoustic-cutoff frequency (71.143 $d^{-1}$) from {Table \ref{tab:dHD156623}} to plot HD 156623 (plotted as a black diamond) in a Kiel diagram alongside known pre-MS and early ZAMS $\delta$ Scutis from \citet{Zwintz14c} in Figure \ref{kiel}. The five hottest pre-MS or early ZAMS $\delta$ Scutis from the sample by \citet{Zwintz14c} -- $\beta$ Pic (A6\,V), HD 261711 (A2\,V), HD 144277 (A1\,V), HD 34282 (A0.5\,V) and HD 37357 (A1\,V) --  have values of their highest excited frequency ranging between 70 $d^{-1}$ and 90 $d^{-1}$. These were plotted in black instead of gray to highlight their similarity to HD 156623, which is the hottest star in the entire sample and its highest excited frequency is within this range. The theoretical blue edge (left dashed line) and the empirical red edge (right dashed line) of the classical $\delta$ Scuti instability strip as defined by \citet{Breger98} were also plotted. The highest excited frequency and HD 156623's placement on the diagram were consistent with the other pre-MS or early ZAMS $\delta$ Scuti stars of the sample as predicted. This star also lies to the left of the theoretical blue edge of the classical instability strip \citep{Breger98}. It was noted that any of the higher candidate frequencies from {Table \ref{tab:2dHD156623}, i.e., 75.211 $d^{-1}$, 89.250 $d^{-1}$ and 93.261 $d^{-1}$}, were also consistent with our interpretation.

\section{Conclusion \label{conclusion}} 

The bRing observatories are the first to identify $\delta$ Scuti pulsations in the young A1\,V star HD 156623, which hosts a gas-rich debris disk.  We detected 16 frequencies in total, 9 of which we confirmed to be real and the remaining 7 we determined to be candidate frequencies. We found that the total amplitude from the 9 confirmed frequencies was small and it is unsurprising that these pulsations were missed in previous studies of the star. We uncovered strong evidence of regularity between the frequencies. We found no evidence of frequency, amplitude, or phase modulation over the course of bRing's observations. The predicted $\log{g}$, along with our new estimated temperature and largest measured confirmed frequency, were used to compare HD 156623 to other pre-MS $\delta$ Scuti stars on a Kiel diagram with theoretical pre-MS and early ZAMS evolutionary tracks. HD 156623 is both consistent with similar stars and lies well within the predicted evolutionary tracks. HD 156623 also lies just beyond the theoretical blue edge of the classical instability strip.

In addition to the frequency analysis, we performed a stellar characterization, spectral, and Sco-Cen membership analysis. We found HD 156623 to be a negligibly reddened $\sim$16\,$\pm$\,7 Myr A1\,V outlying member of the Sco-Cen subgroup Upper-Centaurus Lupus at $d$ $\simeq$ 112 pc. The presence of the gas-rich debris disk supports a young age.

Future work could include a full asteroseismic modeling similar to \citet{Zwintz14} and a more detailed abundance analysis of the star. Before such modeling should be done, however, additional observations from bRing or other instruments should be made to verify the 7 candidate frequencies reported in {Table \ref{tab:2dHD156623}}. These observations should also attempt to discover fainter frequencies. With this information and improved frequencies, this star will provide a better understanding of the young, hot $\delta$ Scutis.

\acknowledgements
SNM is a U.S. Department of Defense SMART scholar sponsored by the U.S. Navy through SSC-LANT.
The results reported herein benefitted from collaborations and/or
information exchange within NASA's Nexus for Exoplanet System Science
(NExSS) research coordination network sponsored by NASA's Science
Mission Directorate.
Part of this research was carried out at the Jet Propulsion
Laboratory, California Institute of Technology, under a contract with
NASA.
TJD and EEM acknowledge support from the Jet Propulsion Laboratory Exoplanetary Science Initiative.
KZ acknowledges support by the Austrian Fonds zur F\"orderung der wissenschaftlichen Forschung (FWF, project V431-NBL) and the Austrian Space Application Programme (ASAP) of the Austrian Research Promotion Agency (FFG).
The authors would like to acknowledge the support staff at both the South African Astronomical Observatory and Siding Spring Observatory for keeping both bRing stations maintained and running.
Construction of the bRing observatory to be sited at Siding Springs, Australia would
not be possible without a University of Rochester University Research Award,
help from Mike Culver and Rich Sarkis (UR), and generous donations of time,
services, and materials from Joe and Debbie Bonvissuto of Freight Expediters,
Michael Akkaoui and his team at Tanury Industries, Robert Harris and Michael
Fay at BCI, Koch Division, Mark Paup, Dave Mellon, and Ray Miller and the
Zippo Tool Room.
This work has made use of data from the European Space Agency (ESA) mission {\it Gaia} (\url{https://www.cosmos.esa.int/gaia}), processed by the {\it Gaia}
Data Processing and Analysis Consortium (DPAC,
\url{https://www.cosmos.esa.int/web/gaia/dpac/consortium}). 
Funding for the DPAC has been provided by national institutions, in particular 
the institutions participating in the {\it Gaia} Multilateral Agreement.
This research was achieved using the POLLUX database
( http://pollux.graal.univ-montp2.fr )
operated at LUPM  (Université Montpellier - CNRS, France
with the support of the PNPS and INSU.

\facilities{bRing-SA, bRing-AU}

\software{Python 3.6.5 \citep{Python}, 
		  scipy \citep{Scipy}, 
          matplotlib \citep{Matplotlib},
          numpy \citep{Numpy},
          astropy \citep{astropy},
          Period04 \citep{Lenz05},
          isoclassify \citep{Huber2017},
          mwdust \citep{Bovy2016}
          }
\bibliography{mellon}{}
\bibliographystyle{aasjournal}

\end{document}